\documentclass{PoS}

\usepackage{ulem,url}
\usepackage{amsmath,amssymb}

\title{Multi-frequency search for Dark Matter: the role of HESS, CTA, and SKA}

\ShortTitle{Dark Matter search with HESS, CTA, and SKA}

\author{\speaker{Geoff Beck}\\
School of Physics, University of the Witwatersrand, Johannesburg, South Africa\\
E-mail: \email{geoffrey.beck@wits.ac.za}}

\author{Sergio Colafrancesco\\
School of Physics, University of the Witwatersrand, Johannesburg, South Africa\\
E-mail: \email{sergio.colafrancesco@wits.ac.za}}

\abstract{
Dark Matter (DM) remains a vital, but elusive, component in our current understanding of the universe. Accordingly, many experimental searches are devoted to uncovering its nature. However, both the existing direct detection methods, and the prominent $\gamma$-ray search with the Fermi Large Area Telescope (Fermi-LAT), are most sensitive to DM particles with masses below 1 TeV, and are significantly less sensitive to the hard spectra produced in annihilation via heavy leptons. The High Energy Stereoscopic System (HESS) has had some success in improving on the Fermi-LAT search for higher mass DM particles, particularly annihilating via heavy lepton states. However, the recent discovery of high J-factor dwarf spheroidal galaxies by the Dark Energy Survey (DES) opens up the possibility of investing more HESS observation time in the search for DM $\gamma$-ray signatures in dwarf galaxies. This work explores the potential of HESS to extend its current limits using these new targets, as well as the future constraints derivable with the up-coming Cherenkov Telescope Array (CTA). These limits are further compared with those we derived at low radio frequencies for the Square Kilometre Array (SKA). Finally, we explore the impact of HESS, CTA, and Fermi-LAT on the phenomenology of the ``Madala" boson hypothesized based on anomalies in the data from the Large Hadron Collider (LHC) run 1. The power of these limits from differing frequency bands is suggestive of a highly effective multi-frequency DM hunt strategy making use of both existing and up-coming Southern African telescopes.
}

\FullConference{
High Energy Astrophysics South Africa Meeting\\
August 25-27, 2016\\
South African Astronomical Observatory\\
Cape Town, South Africa}

\newcommand{\skipthis}[1]{}

\newcommand{\td}[3]{\frac{d^{#3} #1}{d {#2}^{#3}}} 
\renewcommand{\v}[1]{\ensuremath{\mathbf{#1}}} 

\renewcommand{\bar}[1]{\ensuremath{\overline{#1}}}

\begin{document}

\section{Introduction}
\begin{figure}[htbp]
\centering
\includegraphics[scale=0.5]{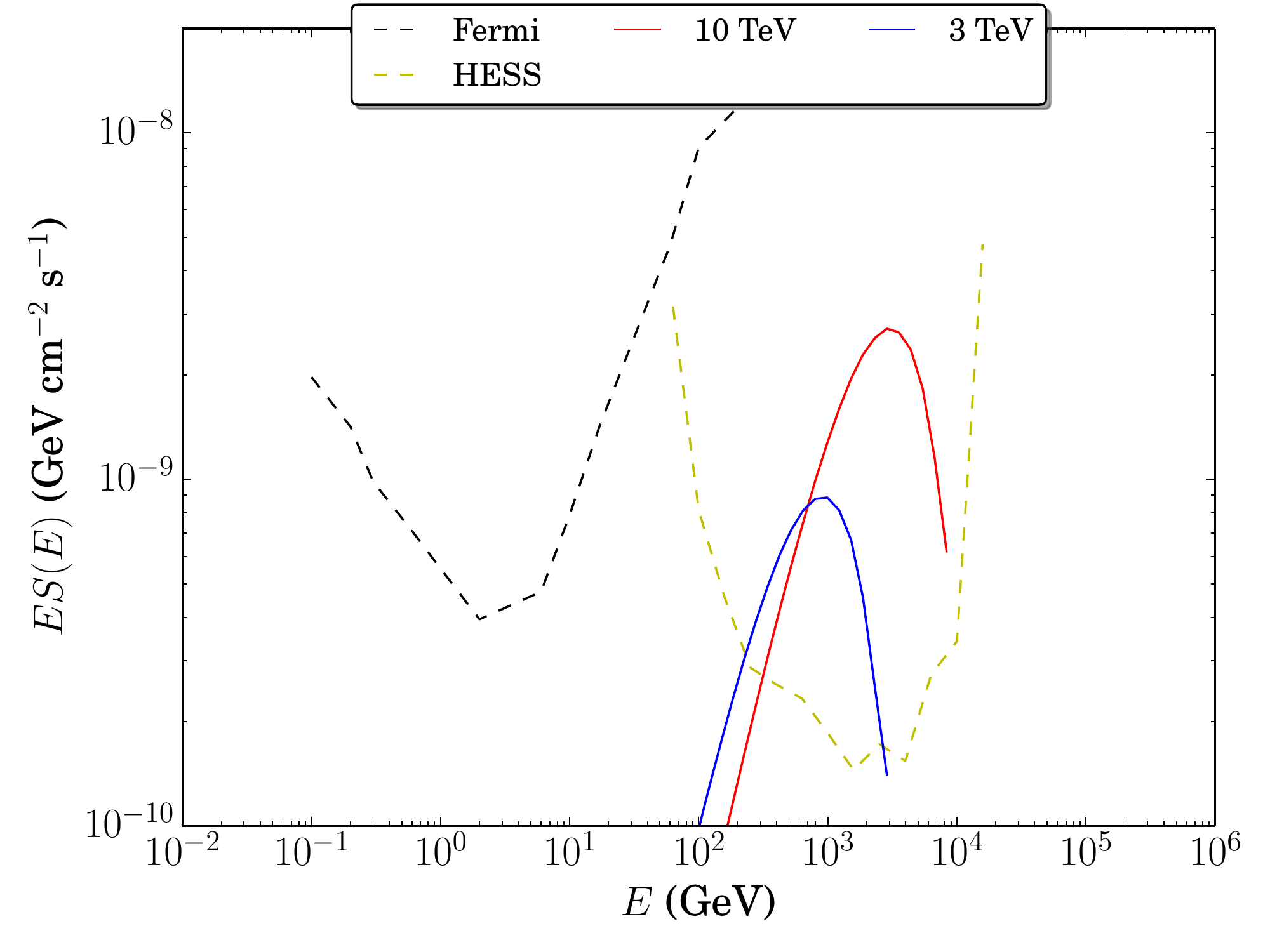}
\caption{Differential $\gamma$-ray spectra from the Reticulum 2 dwarf galaxy for WIMP masses 3 TeV (blue) and 10 TeV (red) annihilating via the $\tau^+\tau^-$ channel. The dashed lines show the Fermi-LAT Pass 8, 10 year sensitivity~\cite{Fermidetails} (black), and the HESS 100 hour differential sensitivity~\cite{hess-perf,funk-cta2013} (yellow). The annihilation cross-section values are chosen simply for the purpose of illustration.}
\label{fig:dsph}
\end{figure}

The nature of dark matter (DM) is one of the primary mysteries in modern physics and cosmology. However, despite the character of DM proving elusive, $\gamma$-ray telescope projects like Fermi-LAT~\cite{fermi-docs} and HESS~\cite{hess-details,hess-perf,funk-cta2013} continue to be effective in probing the parameter space of DM composed of Weakly Interacting Massive Particles (WIMPs)~\cite{hessdwarves2014,Fermidwarves2014,Fermidwarves2015}. Of the two aforementioned projects, HESS has devoted far less time to this search as its sensitivity interval is largely unsuited to the study of the kind of low mass WIMPs that might be consistent with the excess $\gamma$-rays from the galactic centre, as well as the PAMELA~\cite{pamela-docs} and AMS anti-particle experiments~\cite{hooper2014,hooper2011,calore2014,cholis2013}. However, both HESS and the upcoming Cherenkov Telescope Array (CTA)~\cite{cta-docs,funk-cta2013} are admirably suited to probe large mass WIMPs. In this regard we aim to examine how far into the high-mass parameter space these projects can explore, in a manner similar to earlier works on the topic~\cite{doro2012,carr2016}. This being of particular significance as Fermi-LAT is less effective in constraining high-mass WIMPs, particularly those annihilating via $\tau$ lepton states (or other channels that provide harder $\gamma$-ray spectra). This is very effectively illustrated in Figure~\ref{fig:dsph}, which shows that $\gamma$-ray spectra of high-mass WIMPs, annihilating via $\tau$ leptons, falls within the maximum HESS sensitivity region, but is well outside that of Fermi-LAT.

In this work we will make use of the sensitivity calculations from~\cite{hess-perf,funk-cta2013} for HESS and CTA respectively, we note that these may be less accurate than a full mock analysis of the telescope performance.The environments we will use for this study are the Dark Energy Survey (DES) dwarf galaxy candidates~\cite{des,desdwarf} Tucana II, Reticulum II, and the newly discovered Tucana III~\cite{des2015}. This choice of environment is motivated by the extremely large J-factors of these DES dwarf galaxies, which will result in larger DM-induced $\gamma$-ray fluxes than previously studied classical dwarf galaxies. This is of particular importance as it has been recently shown that systematic errors in dwarf galaxy J-factor determination may be considerably under-estimated~\cite{ullio2016}.
The method of analysis will be to produce integrated $\gamma$-ray fluxes for each of the chosen targets; these are then compared to the sensitivity of the HESS and CTA experiments with 100 hours of observation time. As a point of comparison we will also perform the same analysis on the Ophiuchus galaxy cluster, which is visible from the southern hemisphere. Additionally, we will perform this exercise considering both DM fluxes boosted by halo substructure, following the method from \cite{prada2013}, and more conservative cases when such boost effects are absent.\\
In addition to this, we will explore the consequences of DM particle annihilations producing Higgs boson pairs which then subsequently decay. This is based on the recently proposed ``Madala" boson~\cite{madala1}, a 270 GeV particle hypothesised based on multiple anomalies in the LHC run1 data (not subsequently removed by further data). This new boson has Higgs-like couplings within the standard model but also couples to a ``hidden"-sector particle via a mediating scalar. This extra hidden particle is motivated as a candidate for DM as it does not couple to the standard model directly. The mass of the mediating scalar has been tentatively suggested to be greater than 140 - 160 GeV~\cite{madala2,madala3} with the DM mass being half the mass of the mediator. The main decay paths of this mediator are through $W$ and Higgs boson pairs. However, we will focus specifically on the Higgs channel, as it has particularly interesting features. The resulting $\gamma$-ray spectra from the Tucana III dwarf will be compared to the HESS and CTA sensitivities to find null constraints. These will be contrasted with actual Fermi-LAT upper-limits~\cite{Fermidwarves2015} on the Reticulum II dwarf galaxy.

The possible null-constraints from high-energy observations will then be compared to those of the up-coming SKA phase 1~\cite{ska2012}. In order to determine the possible role of southern hemisphere $\gamma$-ray telescopes in a multi-frequency DM search that incorporates up-coming highly sensitive radio experiments like the SKA.
Thus, we will argue that high-energy experiments like HESS and CTA have a particularly important niche role within the hunt for DM by examining how effectively they can constrain the regions of the parameter space least touched by other experiments.\\
This paper is structured as follows: in Sect.~\ref{sec:theory} we detail the models used for halo properties and flux calculations, the results are then presented in Sect.~\ref{sec:results} and final conclusions drawn in Sect.~\ref{sec:conclusion}.

\section{Dark Matter halos and $\gamma$-ray flux}
\label{sec:theory}

The $\gamma$-ray flux produced by a DM halo, integrated over a given energy interval, is written as follows
\begin{equation}
\phi (E_{min},E_{max}, \Delta \Omega, l) = \frac{1}{4\pi} \frac{\langle \sigma V\rangle}{2 m_{\chi}^2} \int_{E_{min}}^{E_{max}} \td{N_{\gamma}}{E_{\gamma}}{} \, d E_{\gamma} \int_{\Delta \Omega}\int_{l} \rho^2 (\v{r}) dl^{\prime}d\Omega^{\prime} \; ,
\end{equation}
where $m_{\chi}$ is the WIMP mass, $\rho$ is the DM halo density profile, $\langle \sigma V\rangle$ is the velocity averaged annihilation cross-section, and $\td{N_{\gamma}}{E_{\gamma}}{}$ is the $\gamma$-ray yield from WIMP annihilations (sourced from PYTHIA~\cite{pythia} routines in DarkSUSY~\cite{darkSUSY}). The astrophysical J-factor encompasses the last two of the above integrals
\begin{equation}
J (\Delta \Omega, l) = \int_{\Delta \Omega}\int_{l} \rho^2 (\v{r}) dl^{\prime}d\Omega^{\prime} \; ,
\end{equation}
with the integral being extended over the line of sight $l$, and $\Delta \Omega$ is the observed solid angle.\\
In this work we will calculate the particle physics factor for an energy interval
\begin{equation}
\psi (> E_{min}) = \frac{1}{4\pi} \frac{\langle \sigma V\rangle}{2 m_{\chi}^2} \int_{E_{min}}^{\infty} \td{N_{\gamma}}{E_{\gamma}}{} \, d E_{\gamma} \; .
\end{equation}
Thus the flux will be found from
\begin{equation}
\phi (> E_{min}) =  \psi (> E_{min}) \times J(\Delta \Omega, l) \; .
\end{equation}
In the case of the Ophiuchus cluster we will calculate the J-factor as follows
\begin{equation}
J_{vir} = 4\pi\int_{0}^{R_{vir}} \rho^2 (r) dr^3 \; ,
\end{equation}
with $R_{vir}$ being the virial radius of the cluster. In order to do this, we use the following DM density profiles
\begin{equation}
\begin{aligned}
\rho_N(r)=\frac{\rho_s}{\frac{r}{r_s}\left(1+\frac{r}{r_s}\right)^{2}} \; ,\\
\rho_{B} (r) = \frac{\rho_s^{\prime}}{\left(1 + \frac{r}{r_s^{\prime}}\right)\left(1+\left(\frac{r}{r_s^{\prime}}\right)^2\right)} \; ,\\
\end{aligned}
\label{eq:nfw}
\end{equation}
where $r_s$ and $r_s^{\prime}$ are the scale radii, $\rho_s$ and $\rho_s^{\prime}$ are the characteristic halo densities, while $\rho_N$ and $\rho_B$ are the NFW and Burkert halo profiles respectively~\cite{nfw1996,Burkert1995}.
For Ophiuchus we use $r_s = 0.611$ Mpc and $\rho_s = 4.31 \times 10^3 \rho_c$ (for NFW) with $\rho_c$ being the critical density. These are found by using a virial radius of $R_{vir} = 2.97$ Mpc and corresponding mass of $M_{vir} = 1.5 \times 10^{15}$ M$_{\odot}$ and using the relations~\cite{ludlow2013}
\begin{equation}
\begin{aligned}
r_s & = \frac{R_{vir}}{c_{vir}} \; , \\
\frac{\rho_s (c_{vir})}{\rho_{c}} & = \frac{\Delta_{c}}{3}\frac{c_{vir}^3}{\ln(1+c_{vir})-\frac{c_{vir}}
{1+c_{vir}}} \; ,
\end{aligned}
\end{equation}
and $c_{vir} (M_{vir})$ is found according to the method described in \cite{prada2012}. The quantity $r_s^{\prime}$ is related to $r_s$ by a factor of $\sim 1.52$, and $\rho_s^{\prime}$ is founded by enforcing a normalization of the integral $4\pi\int_0^{R_{vir}} dr \; \rho_B r^2 = M_{vir}$. The density contrast parameter at collapse $\Delta_c$ is given in flat cosmology by the approximate expression~\cite{bryan1998}
\begin{equation}
\Delta_{c} \approx 18 \pi^2 - 82 x - 39 x^2 \; ,
\end{equation}
with $x = 1.0 - \Omega_m (z)$, where $\Omega_m (z)$ is the matter density parameter at redshift $z$ given by
\begin{equation}
\Omega_m (z)  = \frac{1}{1 + \frac{\Omega_\Lambda (0)}{\Omega_m (0)}(1+z)^{-3}} \; .
\end{equation}

The J-factors found for the targets considered in this work are reported in Table~\ref{tab:jfac}.
\begin{table}[htbp]
\centering
\begin{tabular}{|l|l|l|}
\hline
Halo & J (GeV$^2$ cm$^{-5}$) & Ref\\
\hline
Tucana II & $6.31 \times 10^{18}$ & \cite{Fermidwarves2015} \\
Tucana III & $3.16 \times 10^{19}$ & \cite{des2015}\\
Ret. II & $2.0 \times 10^{19}$ & \cite{bonnivard2015}\\
Ophiuchus (NFW) & $6.54\times 10^{15}$  & - \\
Ophiuchus (Burkert) & $1.85\times 10^{15}$  & - \\
\hline
\end{tabular}
\label{tab:jfac}
\caption{J-factors for each of the studied environment.}
\end{table}
It has been recently shown~\cite{ullio2016} that the systematic errors in the calculation of dwarf galaxy J-factors, introduced through the treatment of the spherical Jeans equation and scaling assumptions for tracers of the inner-halo density, have been greatly under-estimated in the literature. In order to account for this systematic uncertainty, we will show all results with the maximum J-factor uncertainty found in \cite{ullio2016}: this will correspond to a J-factor (and thus flux) reduction by a factor of 4.

Finally, we notice that DM halos may contain substructure, in the form of sub-halos of various masses. These tend to be of a higher concentration than their parent halo and thus enhance the resulting flux from DM annihilation~\cite{pieri2011,Bullock2001}. In order to calculate this amplification (boost) factor we will follow the formulation derived in \cite{prada2013}. The sub-structure boost factor is defined as a luminosity increase caused by integrating over sub-halo luminosities determined by the virial mass and by halo concentration parameters found numerically according to the method discussed in~\cite{prada2012}; we note that a similar method is provided in~\cite{ng2014}. This effect results in a boosting factor (which multiplies the $\gamma$-ray flux) of $b \sim 36$ for Ophiuchus, while dwarf galaxies with masses $\sim 10^7$ M$_{\odot}$ have values $b \sim 3-4$.

\section{Results}
\label{sec:results}

\begin{figure}[htbp]
\centering
\includegraphics[scale=0.36]{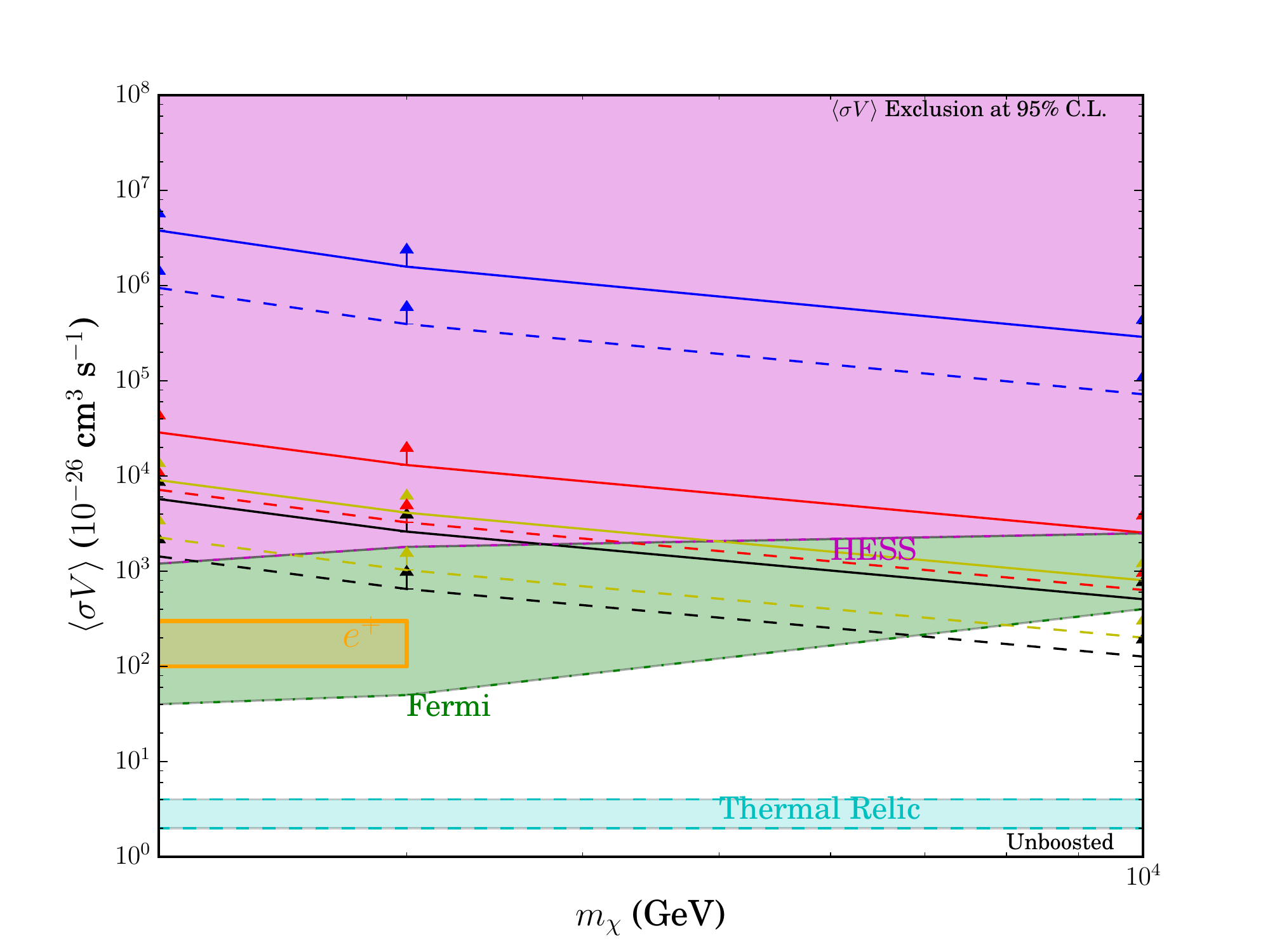}
\includegraphics[scale=0.36]{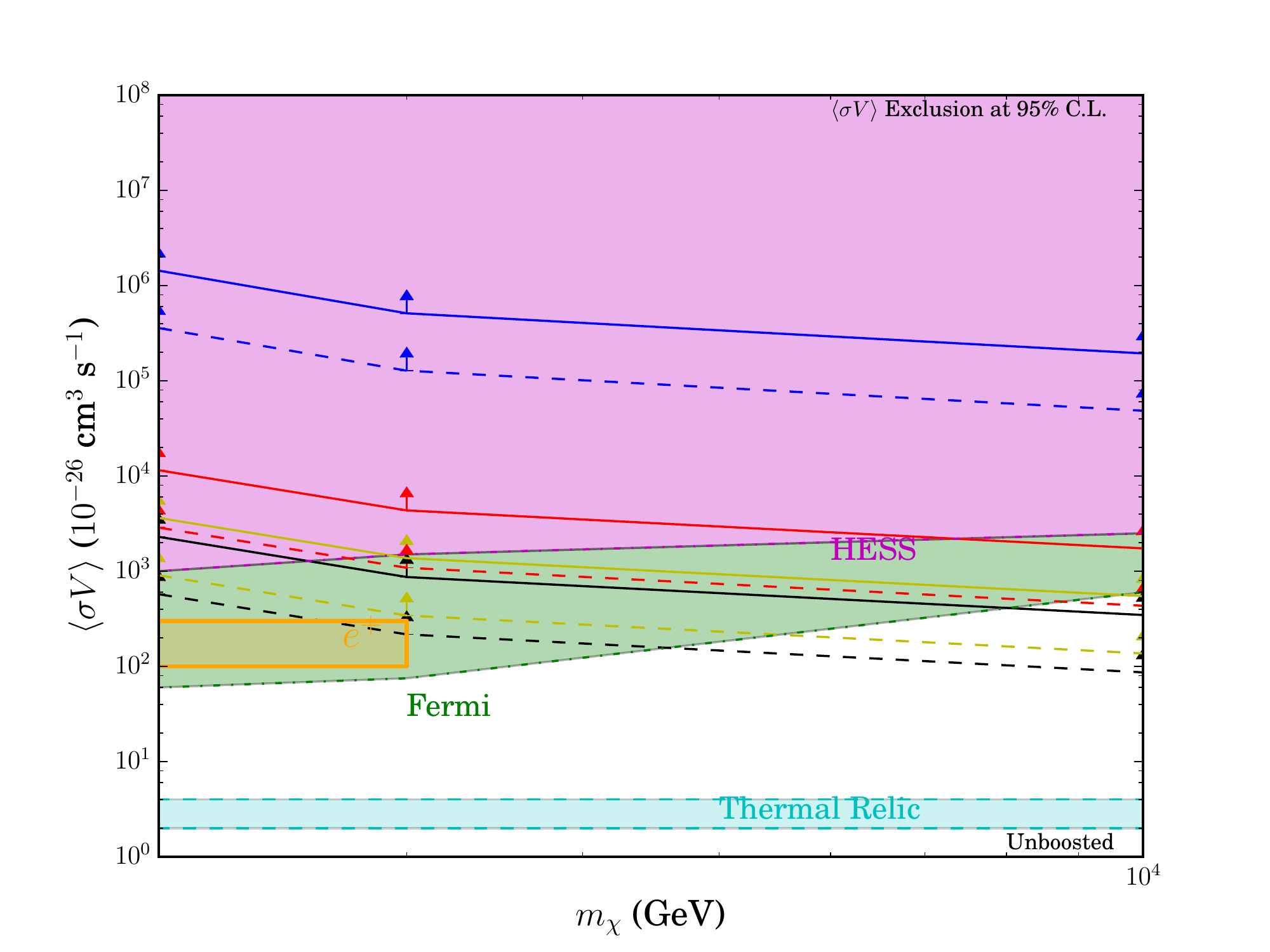}
\includegraphics[scale=0.36]{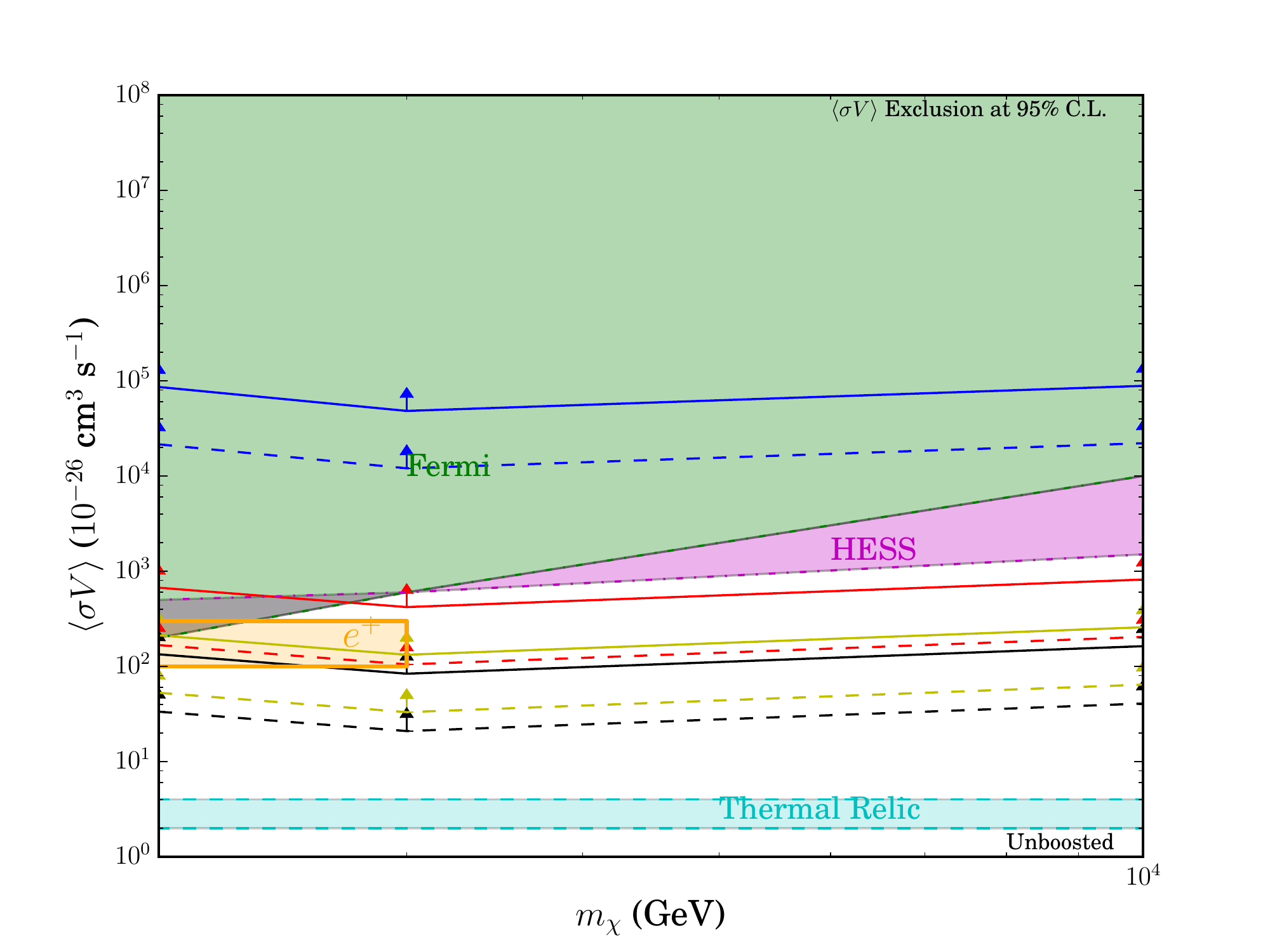}
\caption{Null constraints from 100 hour HESS (solid lines) and CTA (dashed) observations without sub-structure boosting factor of Tucana III (black), Reticulum II (yellow), Tucana II (red), and the Ophiuchus cluster (blue). The arrows show the maximum possible region of uncertainty due to J-factor~\cite{ullio2016}. Existing Fermi-LAT limits are shown in green~\cite{Fermidwarves2015}, and existing HESS limits in pink~\cite{hessdwarves2014}. The AMS positron-excess favoured region is shown in orange~\cite{cholis2013}. Each panel shows a different annihilation channel. Top-Left: annihilation via b-quarks, $b\bar{b}$. Top-Right: annihilation via  $W$-bosons, $W^+W^-$. Bottom: annihilation via $\tau$ leptons, $\tau^+\tau^-$.}
\label{fig:unboost}
\end{figure}

Figure~\ref{fig:unboost} displays the $3\sigma$ null-constraints that can be derived from 100 hours of observation of the targets we consider here via HESS and CTA, comparing these two existing bounds from~\cite{hessdwarves2014,Fermidwarves2015}. In this figure, the case without a substructure boosting effect is shown. The top-left panel demonstrates the sensitivity of the high-energy $\gamma$-ray constraints to the hardness of the target DM spectrum. In this case the $b\bar{b}$ annihilation channel produces a softer spectrum than that of $\tau$ leptons and thus HESS cannot improve on the Fermi-LAT constraints with 100 hours of observation of the chosen targets. CTA can make minor improvements within 100 hours observation of the highest J-factor dwarf galaxies (Tucana III and Reticulum II). The top-right panel shows an improvement in constraints as the $W^+W^-$ channel results in a slight spectral hardening. In this case, both HESS and CTA can improve upon the Fermi-LAT constraints for high-mass WIMPs, above 8 and 5 TeV respectively, with 100 hours of observing time. In the case of CTA this improvement is far more significant than HESS, reaching near an order of magnitude over Fermi-LAT for a 10 TeV WIMP.
The bottom panel shows the most significant result, this being for the case of annihilation via $\tau^+\tau^-$ intermediate states. In this scenario, HESS can achieve nearly an order of magnitude improvement on its previous results between 3 and 10 TeV~\cite{hessdwarves2014} with 100 hours of observation of the high J-factor DES dwarf galaxies, with CTA being able to better this by an additional factor of 2 at all masses shown. Additionally, HESS and CTA can use these targets to probe the parameter space region favoured by AMS positron excess models (shown in orange), which are currently not ruled out in this annihilation channel for both Fermi-LAT and Planck results~\cite{beck2016}. The J-factor uncertainty introduced by the analysis in \cite{ullio2016} does not greatly affect the ability of HESS and CTA to set new limits on the cross-section. It is worth noting that, without any substructure boosting, the dwarf galaxies make far better targets than a galaxy cluster such as Ophiuchus, even before possible background emissions have been considered.

\begin{figure}[htbp]
\centering
\includegraphics[scale=0.36]{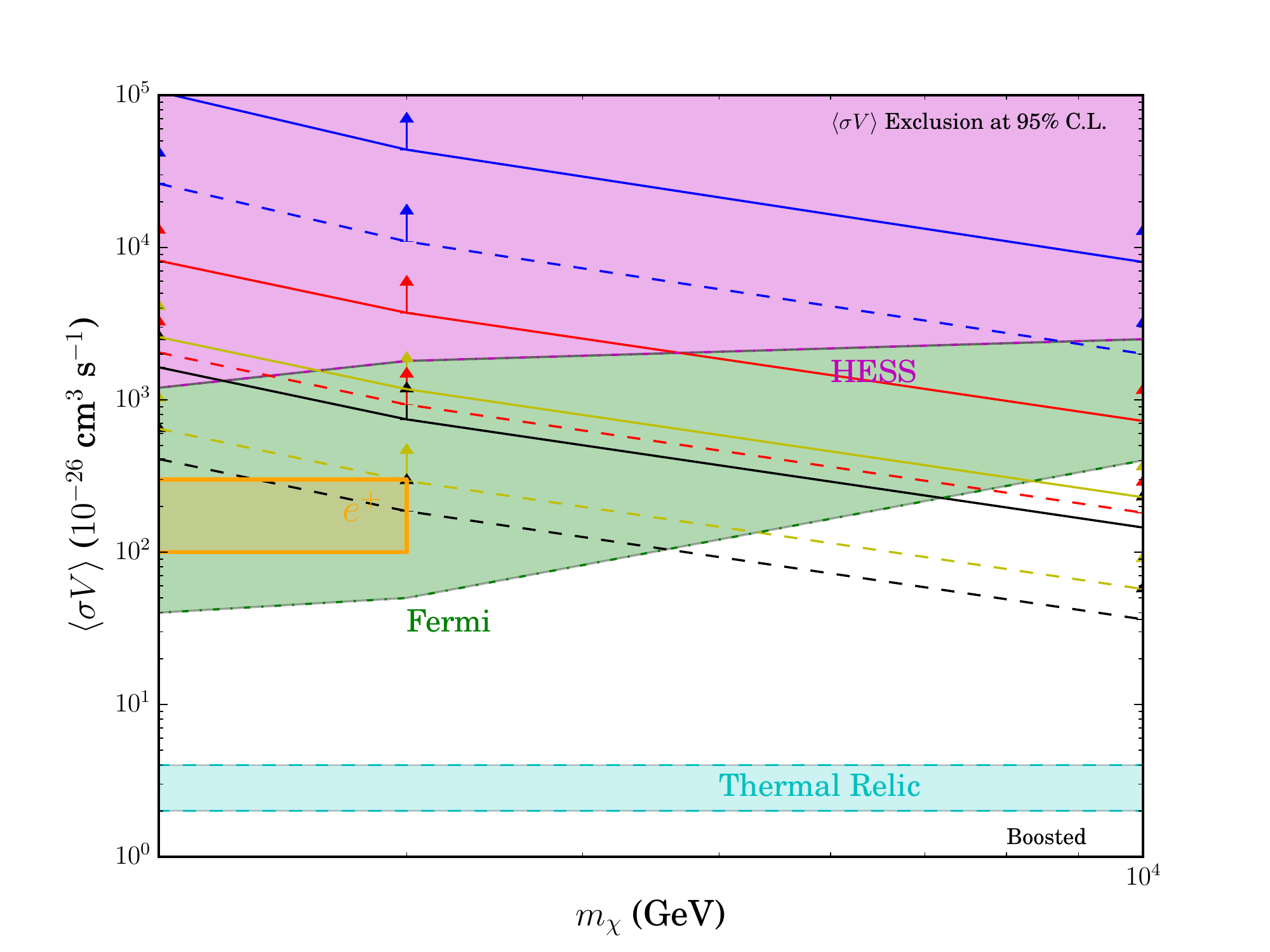}
\includegraphics[scale=0.36]{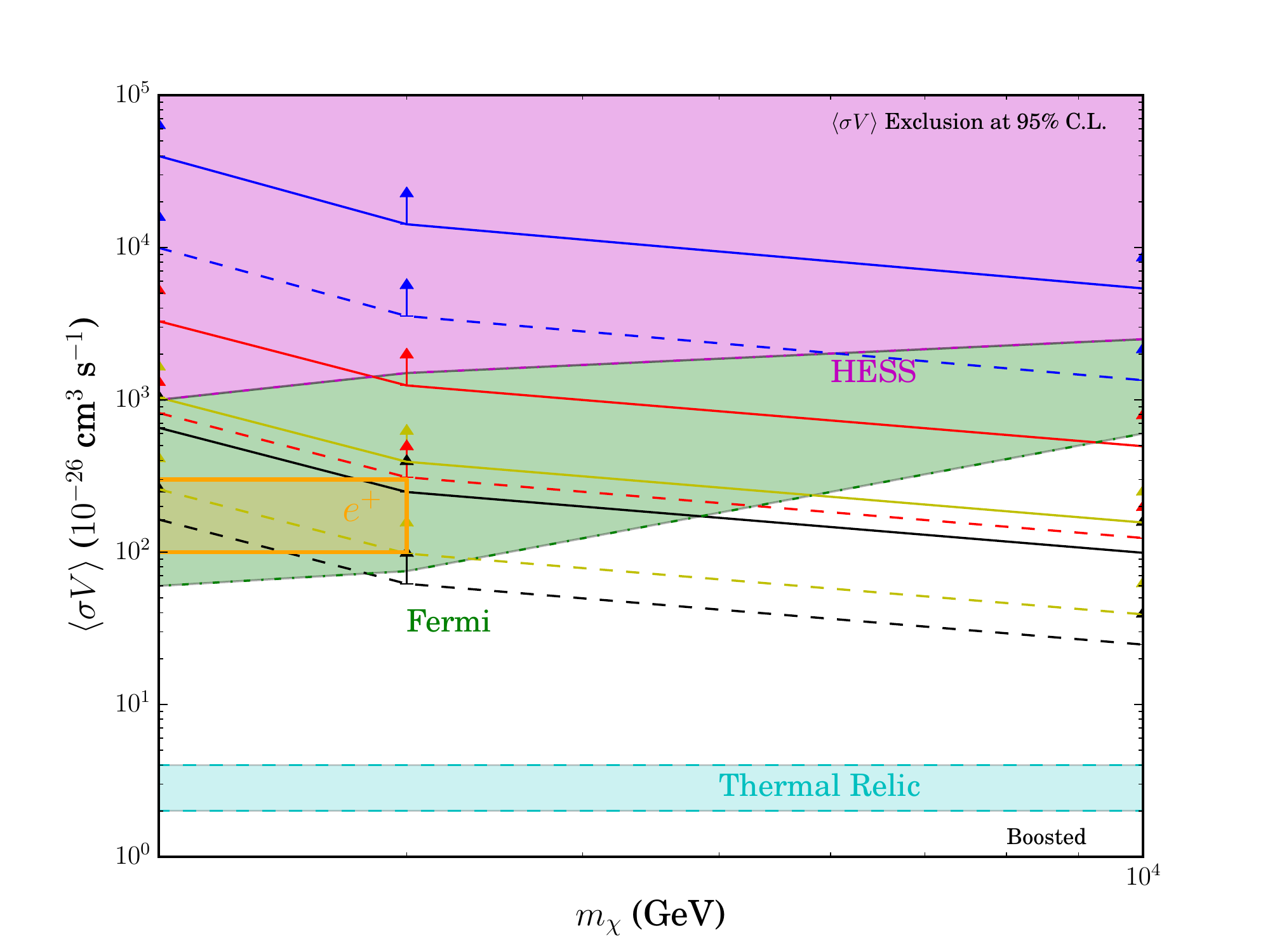}
\includegraphics[scale=0.36]{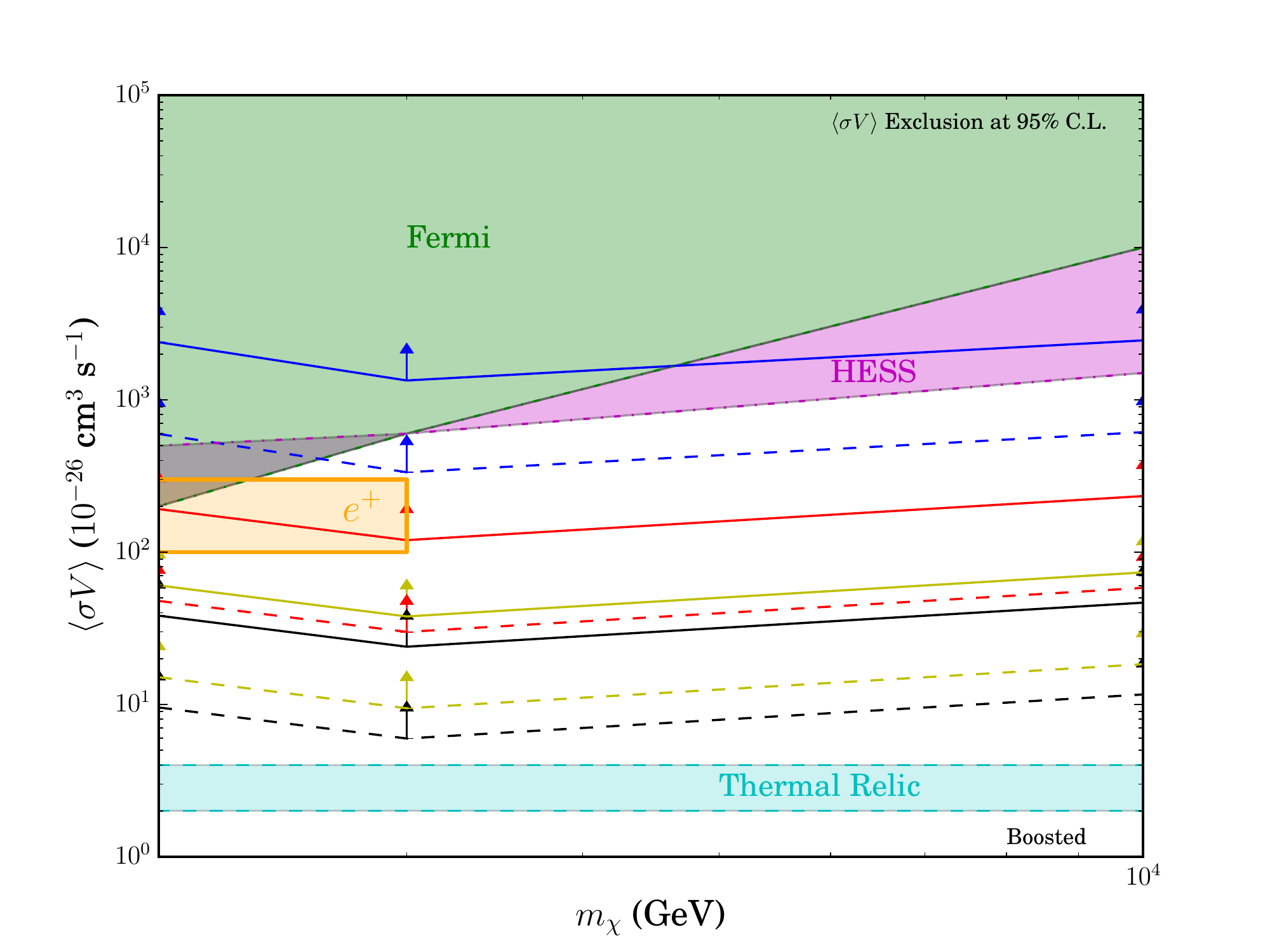}
\caption{Null constraints from 100 hour HESS (solid lines) and CTA (dashed) observations including sub-structure boosting factor of Tucana III (black), Reticulum II (yellow), Tucana II (red), and the Ophiuchus cluster (blue). The arrows show the maximum possible region of uncertainty due to J-factor~\cite{ullio2016}. Fermi-LAT limits are shown in green~\cite{Fermidwarves2015}, HESS limits in pink~\cite{hessdwarves2014}. The AMS positron-excess favoured region is shown in orange~\cite{cholis2013}. Each panel shows a different annihilation channel. Top-Left: annihilation via b-quarks, $b\bar{b}$. Top-Right: annihilation via  $W$-bosons, $W^+W^-$. Bottom: annihilation via $\tau$ leptons, $\tau^+\tau^-$.}
\label{fig:boost}
\end{figure}

Figure~\ref{fig:boost} shows the $3\sigma$ null-constraints that can be derived through 100 hours of observation via HESS and CTA, comparing these two existing bounds from~\cite{hessdwarves2014,Fermidwarves2015}. In this figure, the case including the substructure boosting effect is shown. As expected, given the dwarf boosting factor of $\sim 4$, the potential null-constraints are substantially improved, to the point where even the $b\bar{b}$ channel provides a candidate for the extension of the current HESS limits within 100 hours of observing time. Of particular note is that a galaxy cluster like Ophiuchus is still an unviable target for HESS, despite its larger boosting factor of $\sim 36$. For CTA, Ophiuchus can provide marginal improvements in the $\tau$ lepton channel but these are likely insignificant should there be any source of background $\gamma$-ray fluxes from non-DM sources in the cluster. As a  coincidence it is also worth noting that the dwarf J-factor maximum uncertainty from \cite{ullio2016} is very similar to the boost gained from the substructure effect in dwarf galaxies around $10^7$ $M_{\odot}$.

\begin{figure}[htbp]
\centering
\includegraphics[scale=0.36]{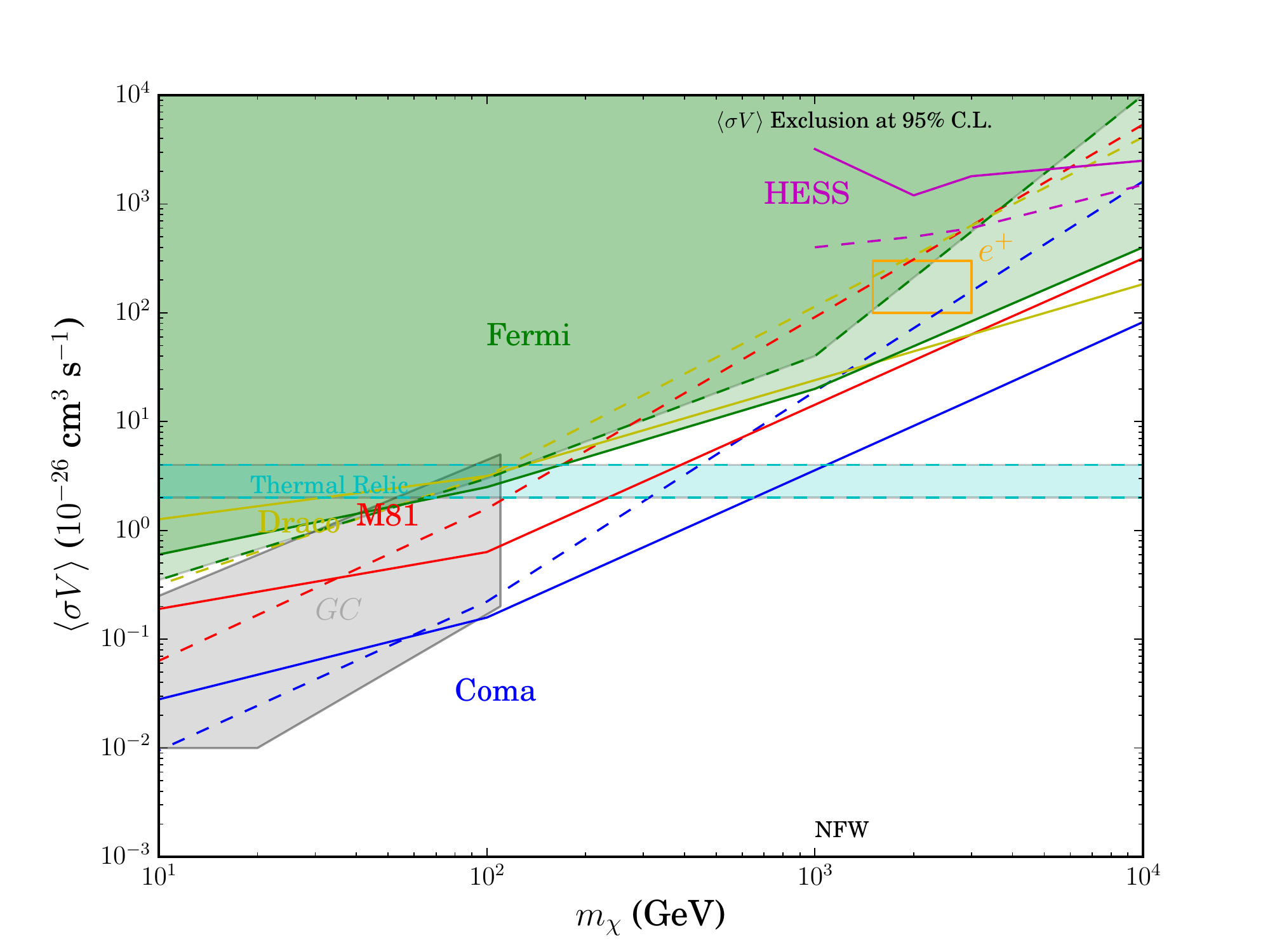}
\includegraphics[scale=0.36]{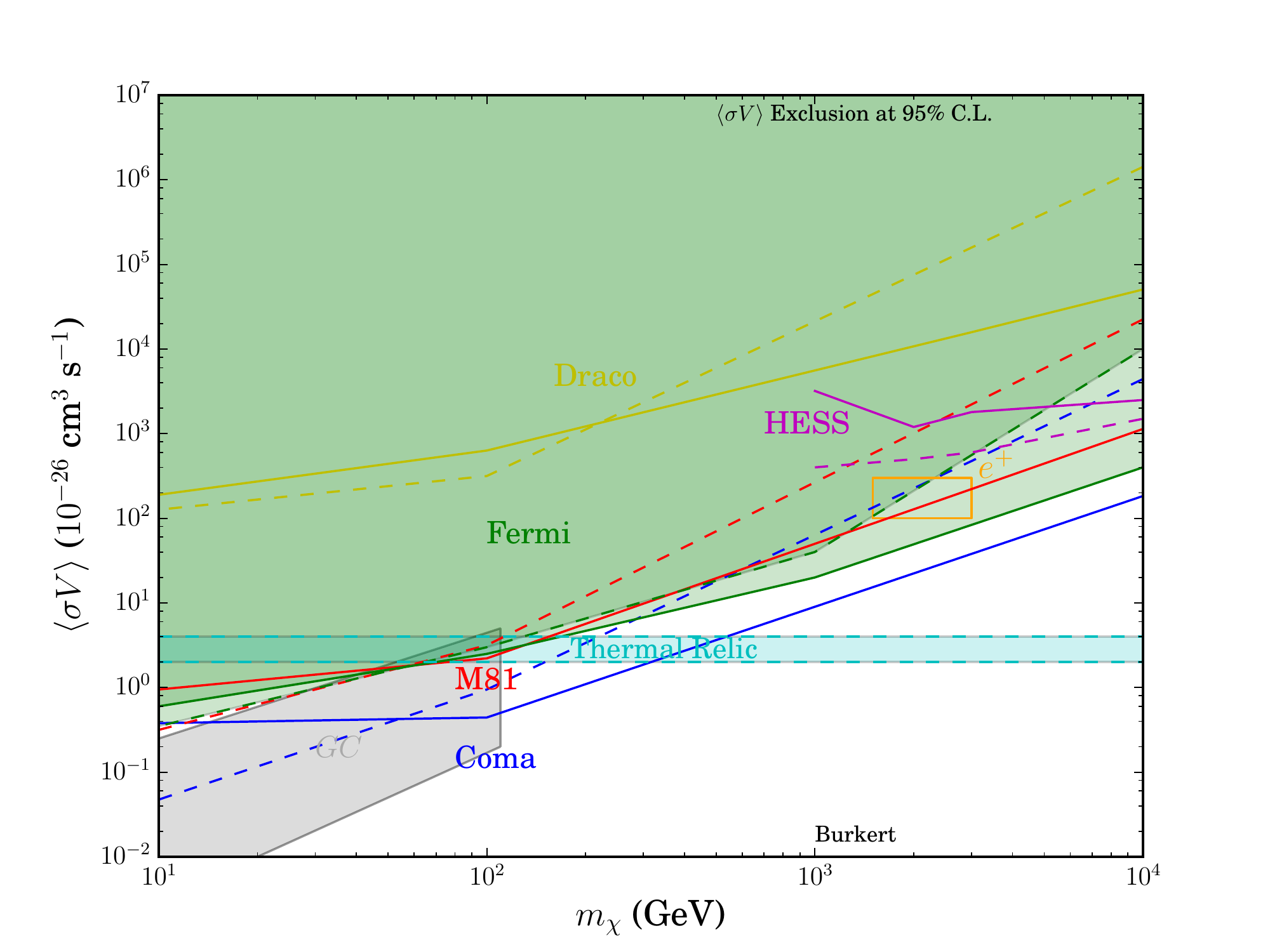}
\caption{The SKA phase-1 null constraints from 100 hour observations. Projected constraints (including effect of dominant cosmic-ray backgrounds) shown for the Coma cluster (blue), M81 galaxy (red), and Draco dwarf galaxy (yellow). Fermi-LAT limits are shown in green~\cite{Fermidwarves2015}, HESS limits in pink~\cite{hessdwarves2014}. The grey region shows the region of galactic-centre best fit~\cite{calore2014}, while orange shows the AMS positron excess region~\cite{cholis2013}. Figure from \cite{beck2016}.}
\label{fig:ska}
\end{figure}

In Figure~\ref{fig:ska} we display the potential constraints derivable with the SKA from DM-induced synchrotron radiation within reference targets like the Draco dwarf galaxy, the M81 galaxy, and the Coma cluster, as considered in a recent study by \cite{beck2016} (see discussion and references therein for full details). These constraints were derived by assuming a power-law background spectrum in each target, normalized using the available data. The minimal value of the annihilation cross-section is then found, such that the DM synchrotron emission profile can be extracted from the background given the capabilities of the SKA. The weakness of the Draco results can be largely attributed to the available data only observing small region in the centre of the dwarf galaxy. The great power of the SKA to probe the WIMP parameter space with just 100 hours of observation is clearly evident in these results. However, the curves of particular interest are those for $\tau$ lepton annihilations (dashed lines). It is clear that, in both the NFW and Burkert density profile cases, these potential constraints are substantially weaker than their $b$-quark counterparts. Of particular importance is the fact that these constraints become either weaker than the existing HESS results above 3 TeV, or very similar to them. Thus, the potential constraints from HESS and CTA, displayed in Figs~\ref{fig:unboost} and \ref{fig:boost}, can play a role in expanding our ability to constrain the WIMP parameter space, thus acting as an important supplement to experiments like SKA and Fermi-LAT that are more sensitive to softer spectra.

\begin{figure}[htbp]
\centering
\includegraphics[scale=0.5]{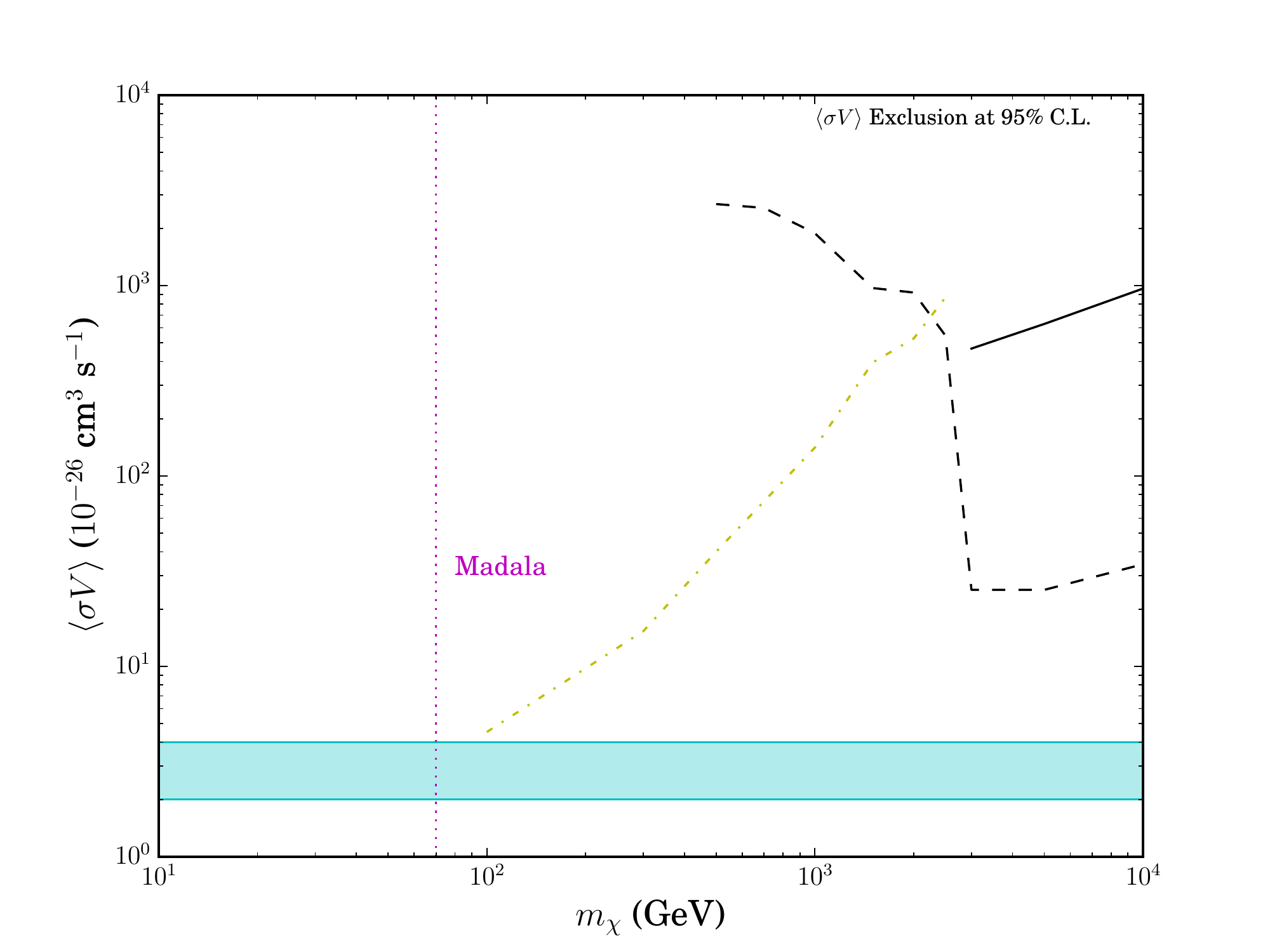}
\caption{Potential null constraints on $\langle \sigma V\rangle$ for the Higgs boson annihilation channel from 100 hours observation from Tucana III. Those from HESS are shown as the solid line, and from CTA as the dashed line. The yellow dot-dashed line shows the constraints derived from Fermi-LAT upper-limits on Reticulum II emissions~\cite{Fermidwarves2015}. The vertical magenta line displays the lower bound of 70 GeV on the mass of DM for the Madala hypothesis~\cite{madala2,madala3}. Limits are derived from differential sensitivities~\cite{funk-cta2013,Fermidetails}.}
\label{fig:higgs}
\end{figure}

Finally, the consequences of DM annihilation via Higgs bosons are shown in Figure~\ref{fig:higgs}. The $\gamma$-ray yield data~\cite{ppdmcb1,ppdmcb2} for this channel show a large increase at the mass of the DM particle, resulting in a line-like feature in the emission spectrum from a DM halo. The maximum of this feature was used to constrain the cross-sections that would be allowed by 100 hour null constraints on the Tucana III dwarf galaxy. This results in very ``spikey" constraint curves, as it is highly sensitive to the shape of the experiment's differential sensitivity profile. The constraints from Fermi-LAT data on Reticulum II~\cite{Fermidwarves2015} are also shown for comparison. It is clear that Fermi-LAT can probe the low mass spectrum more effectively, while CTA becomes competitive soon after the lower bound on the DM particle from the Madala hypothesis and displays great potential to produce constraints well below the thermal relic limit on particles with both intermediate and very large masses, these being more severe than even the best case scenario with heavy lepton annihilations. HESS shows a similar ability to put stringent constraints on this annihilation channel with 100 hours of observation, being able to probe down to the thermal relic level even for particles of mass 1 TeV and higher.

\section{Conclusions}
\label{sec:conclusion}

This work has shown that there is great potential for the current iteration of HESS to extend the Fermi-LAT constraints on high-mass WIMPS, particularly in the region of the parameter space that boasts the weakest constraints. This is found to be independent of the use of a substructure boosting factor, although its presence provides stronger constraints. The up-coming CTA experiment will have even greater ability to probe the high-mass region of the WIMP parameter space, additionally being more sensitive to the softer $W$-boson and $b$-quark annihilation channels. Of particular note is the fact that CTA can improve on Fermi-LAT by nearly an order of magnitude for WIMP masses between 3 and 10 TeV in the $\tau$ lepton annihilation channel. The dwarf galaxies Tucana II \& III as well as Reticulum II are highlighted as being of particular interest to the continuing search for DM due to their exceptionally large J-factors. Furthermore, we demonstrated that HESS and CTA DM searches can play a role in supplementing future experiments with high DM sensitivity like the SKA, which are very sensitive to lower WIMP masses~\cite{gsp2015,Colafrancesco2015,beck2016}, but also place weaker constraints on high-mass WIMPs with hard annihilation spectra. Finally, we showed that very powerful constraints, well below Fermi-LAT above 100 GeV masses, can be derived using HESS and CTA for WIMPs associated with the recently hypothesized ``Madala" boson which was motivated by anomalies in LHC run1 data. This means that HESS and CTA boast the potential to rule out the Madala-associated candidate particle as a major component of DM, provided its annihilations predominantly result in the production of Higgs boson pairs.\\
These arguments fully motivate a multi-frequency search, incorporating Southern-African experiments, with the potential to greatly advance the search for WIMP DM across the mass spectrum. This is of particular significance as the radio and $\gamma$-ray observations are sensitive to differing astrophysical uncertainties and cosmic backgrounds.


\begin{thebibliography}{99}

\bibitem{fermi-docs}
Atwood, W. B. \textit{et al}. for the Fermi/LAT collaboration, 2009, Astrophys. J., \textbf{697},1071, arXiv:0902.1089 [astro-ph].

\bibitem{hess-details}
\url{https://www.mpi-hd.mpg.de/hfm/HESS/}

\bibitem{hess-perf}
HESS Collaboration, Aharonian, F. \textit{et al}., 2006, Astron.Astrophys., \textbf{457}, 899.

\bibitem{funk-cta2013}
Funk, S. \& Hinton, J., 2013, APh, \textbf{43}, 348.

\bibitem{hessdwarves2014}
Abramowski, A. \textit{et al}. for the HESS colloboration, 2014, arXiv: 1410.2589 [astro-ph].

\bibitem{Fermidwarves2014}
Ackermann, M. et al. for the Fermi-LAT collaboration, 2014, Phys. Rev. D, \textbf{89}, 042001, arXiv:1310.0828 [astro-ph.HE].

\bibitem{Fermidwarves2015}
Drlica-Wagner, A. \textit{et al}. for the Fermi-LAT collaboration \& Abbott, T. \textit{et al}. for the DES collaboration, 2015, arXiv: 1503.02632 [astro-ph].

\bibitem{pamela-docs}
Picozza, P. \textit{et al}., 2007, Astropart. Phys., \textbf{27}(4), 296.

\bibitem{hooper2014}
Hooper, D., Linden, T. \& Mertsch, P., 2015, JCAP, \textbf{03}, 021, arXiv:1410.1527 [astro-ph].

\bibitem{hooper2011}
Hooper, D. \& Linden, T., 2011, Phys. Rev. D, \textbf{84}, 123005.

\bibitem{calore2014}
Calore, F., Cholis, I., McCabe, C. \& Weniger, C., 2015, Phys. Rev. D, \textbf{91}, 063003, arXiv:1411.4647 [astro-ph].

\bibitem{cholis2013}
Cholis, I. \& Hooper, D., 2013, Phys. Rev. D, \textbf{88}, 023013, arXiv:1304.1840 [astro-ph].

\bibitem{cta-docs}
\url{https://portal.cta-observatory.org/Pages/Home.aspx}

\bibitem{doro2012}
Doro, M. \textit{et al}., 2013, Astroparticle Physics, 43, 189.

\bibitem{carr2016}
Carr, J. \textit{et al}., 2015, In Proceedings of the 34th International Cosmic Ray Conference (ICRC2015), arXiv: 1508.06128.

\bibitem{Fermidetails}
\url{http://www.slac.stanford.edu/exp/glast/groups/canda/lat_Performance.htm}


\bibitem{des}
\url{http://www.darkenergysurvey.org}

\bibitem{desdwarf}
Bechtol, K. \textit{et al}. (The DES Collaboration), Accepted toApJ (2015), arXiv:1503.02584 [astro-ph.GA].

\bibitem{des2015}
Drlica-Wagner, A. \textit{et al}. (DES), 2015, Astrophys. J., \textbf{813}, 109, arXiv: 1508.03622.

\bibitem{ullio2016}
Ullio, P. \& Mauro, V., JCAP, 2016, \textbf{07}, 025.

\bibitem{prada2013}
Sanchez-Conde, M. \& Prada, F., MNRAS, 2014, \textbf{442} (3), 2271, arXiv:1312.1729 [astro-ph].

\bibitem{madala1}
von Buddenbrock, S. \text{et al}., 2015, arXiv:1506.00612 [hep-ph].

\bibitem{madala2}
von Buddenbrock, S. \textit{et al}., 2016, arXiv:1606.01674 [hep-ph]

\bibitem{madala3}
Mellado, B., 2016, presentation at University of the Witwatersrand.

\bibitem{ska2012}
Dewdney, P., Turner, W., Millenaar, R., McCool, R., Lazio, J. \& Cornwell, T., SKA baseline design document, 2012, \url{http://www.skatelescope.org/wp-content/uploads/2012/07/SKA-TEL-SKO-DD-001-1_BaselineDesign1.pdf}

\bibitem{pythia}
Sj\"ostrand, T., 1994, Comput. Phys. Commun., 82, 74.

\bibitem{darkSUSY}
Gondolo, P., Edsjo, J., Ullio, P., {\it et al.}., 2004, JCAP, \textbf{0407}, 008.

\bibitem{nfw1996}
Navarro, J. F., Frenk, C. S. \& White, S. D. M., 1996, ApJ, \textbf{462}, 563.

\bibitem{Burkert1995}
Burkert, A., 1995, ApJ, \textbf{447}, L25.

\bibitem{ludlow2013}
Ludlow, A. D. {\it et al.}, 2013, MNRAS, \textbf{432}, 1103L.

\bibitem{prada2012}
Prada, F. \textit{et al}., 2012, MNRAS, \textbf{423} (4), 3018, arXiv:1104.5130 [astro-ph].

\bibitem{bryan1998}
Bryan, G. \& Norman M., 1998, ApJ, 495, 80

\bibitem{bonnivard2015}
Bonnivard, V. \textit{et al}., 2015, ApJ, \textbf{808}, L36.

\bibitem{pieri2011}
Pieri L. {\it et al.}, 2011, Phys. Rev. D., \textbf{83}, 023518.

\bibitem{Bullock2001}
Bullock, J.S. {\it et al.}., 2001, MNRAS, 321, 559

\bibitem{ng2014}
Ng, K. \textit{et al}., 2014, Phys. Rev. D, \textbf{89}, 083001, arXiv: 1310.1915 [astro-ph.CO].

\bibitem{beck2016}
Beck, G. \& Colafrancesco, S., 2016, JCAP, \textbf{05}, 013.

\bibitem{ppdmcb1}
Cirelli, M., \textit{et al}.,2011, JCAP, 1103, 051. Erratum: 2012, JCAP, 1210, E01,  arXiv 1012.4515.

\bibitem{ppdmcb2}
Ciafaloni, P. \textit{et al}., 2011, JCAP 1103, 019, arXiv 1009.0224.

\bibitem{gsp2015}
Colafrancesco, S., Marchegiani, P. \& Beck, G., 2015, JCAP, \textbf{02}, 032C.

\bibitem{Colafrancesco2015}
Colafrancesco , S. et al, 2015, \textit{Probing the nature of dark matter with the SKA}, Proceedings of Science: \textit{Advancing Astrophysics with the SKA}, arXiv:1502.03738 [astro-ph].



\end{thebibliography}
\end{document}